\documentclass[12pt,preprint]{aastex}
\begin{document}                                                                     

\title{Massive Star Multiplicity: The Cepheid W Sgr
  \altaffilmark{1} }


\author{Nancy Remage Evans }
\affil{Smithsonian Astropnysical Observatory,  MS 4,   
 60 Garden St., Cambridge, MA 02138}

\author{Derck Massa }     
\affil{NASA's GSFC, SGT Inc.}

\author{Charles Proffitt }     
\affil{Science Programs, Computer Sciences Corporation,            
Space Telescope Science Institute, 3700 San Martin Dr. Baltimore, MD 21218, US}

\email{nevans@cfa.harvard.edu}


\altaffiltext{1}{Based on observations made with the NASA/ESA Hubble Space                
             Telescope, obtained at the Space Telescope Science                       
             Institute, which is operated by the Association of                       
             Universities for Research in Astronomy, Inc. under NASA Contract NAS5-26555. }


\begin{abstract}

We have obtained spectra of the W Sgr system with the STIS 
spectrograph on the Hubble Space Telescope.  The spectra resolve
the system into a distant companion B which is the hottest star in 
the system  and the spectroscopic binary (A = Aa + Ab). A and
B are separated by $0\farcs16$.  We have extracted 
the spectra of both of these.  We see no flux in the Aa + Ab spectrum 
which cannot be accounted for by the Cepheid, and put an upper limit on 
the spectral type and mass of the companion Ab of F5 V and $\leq$1.4M$\sun$.   Using the 
orbit from HST FGS measurements from Benedict, et al., this results in 
an upper limit to the mass of the Cepheid of $\leq$5.4M$\sun$.  We also discuss 
two possible distant companions.  Based on photometry from the 2MASS
Point Source Catalog, they are not physical companions of the W Sgr 
system.  

\end{abstract}



%% In the first two sections, notice the use of the natbibcitep

\section{Introduction}

Because W Sgr = HD 164975 = ADS 1102 is a very bright Cepheid, its 
radial velocities have been measured for more than a century.  With hindsight, we can even
say that the system has been teasing us for that long.  

Although there were suggestions that there might be orbital 
motion as well as pulsation in the velocities, 
it was not until 1989 (Babel, et al. 1989) that a spectroscopic orbit 
was determined from all the previous velocity observations.  
This was achieved 
largely because of the precision of the CORAVEL velocities, since  
the orbital velocity amplitude K is only 2.4 km sec$^{-1}$. 
  Further radial velocity work (Albrow and Cottrell, 1996; 
Pettersen, Cottrell, and Albrow, 2004) has continued to improve
the orbit.  

 In the meantime, the system had apparently been resolved with speckle 
interferometry (Morgan 1978) with a separation between the components of 
$0\farcs116$.   It was, however, not resolved by Bonneau, et al. (1980).
Babel, et al. realized that  their 4.9 year orbit
was not compatible with the interferometric companion.

Ultraviolet low resolution                          
 International Ultraviolet Explorer (IUE) 
spectra  showed that a hot companion was clearly present  
(B\"ohm-Vitense, 1985; B\"ohm-Vitense and Proffitt, 1985; 
Evans, 1991).

A breakthrough came with the HST Fine Guidance Sensor
(FGS) observations of Benedict, et al. (2007),
whose main aim was to measure the stellar parallax.
They were able to measure the orbital motion of the primary (Cepheid)
on the sky with the FGS.  Combining these measures with the spectroscopic orbit of 
Petterson, Cottrell, and Albrow, they were able to solve for the orbital
parameters including the inclination.  By assuming a mass of the 
companion inferred from a spectral type, a mass for the Cepheid can 
be derived from the orbit.

Determining information about the companion in the 4.3 year orbit is the 
topic of this study.  It can be noted from the summary above that there 
have been some contradictions and challenges in obtaining information 
about the orbit.  The low orbital velocity amplitude necessitates 
high accuracy velocities, and additional data would be valuable.
Information on resolution of the system has  not been consistent.  
In addition, using a mass of the secondary inferred from the IUE 
spectral type and the mass function of the orbit leads to a very 
small inclination for a reasonable Cepheid mass.

W Sgr = ADS 11029 also has two possible faint companions at 33" and 47" separations,
which will also be discussed here to investigate the 
membership of the system.
Wide companions are particularly useful in that they contain information
about interactions the system has had with other stars in the 
vicinity.  Possible companions such as these are worth 
investigating because they are relatively low mass stars.  
Patience, et al. (2002) find that high mass systems are more likely 
to have small mass ratios than low mass stars.  Furthermore, 
higher mass systems have larger maximum separations of companions
(Kraus and Hillenbrand, 2007a).  It is valuable to investigate 
further both 
these characteristics for stars as massive as Cepheids (typically
about 5 M$\sun$).  Differences between high mass systems and 
low mass systems point to  differences in formation scenarios. 
 
Assembling information about the properties of binary and multiple systems 
is a key ingredient in studying star formation processes.  Currently it is 
not clear how capture and fragmentation processes are balanced, nor the 
relation between multiple systems and disks. Furthermore, once the system configuration
has been set, dynamical evolution can produce additional changes.  It is becoming
clear, however, that high mass systems have a higher binary frequency 
and larger maximum separation than lower mass systems.  

\section{Observations and Data Reduction}

A number of images of the W Sgr system were taken under 
HST programs (D. Massa, P. I.).  The system was 
 on the target list of a program to resolve 
Cepheids with composite spectra and ultimately 
determine the distances (see 
Massa and Evans, 2008).  W Sgr was a good candidate
for resolution because the low orbital velocity 
amplitude made a low inclination likely.

\subsection{STIS Image}

A series of Space Telescope Imaging Spectrograph (STIS)
spectra were obtained at changing roll angles, and 
at different times in the orbital period. The observation
program was very similar to that of AW Per (Massa and Evans, 2008).
That paper provides a full discussion of the observing details.   
The images were made with the G230L grating and the 
STIS NUV-MAMA detector, which provided spectral
coverage from 1800 to 3400 \AA.  In order to determine 
the separation (including the orientation), observations
were taken at 3 roll angles.  At each of these, observations were
made with the target at 3 locations on the detector offset by $\pm$$0\farcs1$.

The STIS image discussed below is shown in Fig. 1.  
The answer to some of the puzzles about
the system is immediately obvious in the image.   The Cepheid dominates 
the spectrum at the longest wavelengths.  Using reasonable estimates for the 
mass of the Cepheid and the companion, the spectroscopic system at the 
distance of the Cepheid would only be a few $0\farcs01$, and hence unresolved 
from the Cepheid on the STIS image.  The hottest star in the system,
the star which dominates the IUE spectrum from 1200 to 1800 \AA, however,
is clearly resolved from the Cepheid.  We designate the stars in the image
Aa: the Cepheid, Ab: the companion in the spectroscopic orbit, and B: 
the resolved (hottest) star in the STIS image. 
The purpose of this study is to  determine the properties of Ab.

\subsection{Measuring the Separation}

The extraction of spectra from the image requires the determination of a point spread
function (PSF) of the spectrum, which varies with position on the detector.  It is
evident in Fig. 1 that the two spectra are close enough that a well understood
PSF is needed for the extraction. 
The native 1024x1024 format of the STIS NUV MAMA detector anode array when used with
the  G230L grating has a plate scale of about $0\farcs0248$/pixel in the cross dispersion
direction. In the near-UV this provides only about one pixel per resolution element,
and so is rather undersampled.
There is however, a way to measure the subsampled PSF which is needed to 
extract the two W Sgr spectra.  The G230L spectrum is
rotated at a slight angle with respect to the anode array which measures the locations
of photons incident on the MAMA detector. From one end to the other, the spectrum
crosses about 16 rows of pixels, effectively subsampling the PSF in the
cross-dispersion direction. The relative location of the spectrum in the cross dispersion
direction as a function of wavelength will be referred to as the spectral trace.

To measure the PSF we use an observation of a single star taken with the same mode as the
science observations. The W Sgr G230L observations used the F25NDQ1 neutral density
filter to keep the target below allowed bright object limits.  However, for this
filter/aperture combination,  only a single calibration observation was ever 
done. This was data set o3zx08k0, with the WD standard GD153 as the target. For the W
Sgr observations we used this observation to provide our reference cross-dispersion
PSF.  

The subsampled PSFs were prepared by using the STSDAS STIS package task ``wx2d" to
subsample each row of the original flatfielded images of the PSF stars by 8x in the
cross dispersion direction. The resulting cross-dispersion profile in each column was
then smoothed over 32 pixels in the dispersion direction to average over the different
pixel centerings and produce our adopted PSF profile as a function of wavelength.  
The fitting
of these PSFs  also allowed us to measure the tilt of the spectral trace as a
function of wavelength over the detector.

As an example of fitting the components and projected separations,
we will use observation o6f109020 of W Sgr.  The flat fielded image is shown in Fig 1.
For each flat fielded science image of the target star, we also used the wx2d task to
produce images that were sub-sampled by 8x in the cross dispersion direction.  Then
at each pixel in the dispersion direction we fit the observed profile as a sum of
multiple PSFs using a Marquardt-Levenson algorithm, and allowing the location and size
of each component to be free parameters. This fit yields a position and a relative
count rate for each component as a function of wavelength.

The location of the spectral trace is not completely stable 
from observation-to-observation. We adjusted
the trace as initially determined from the PSF star to subtract 
any systematic shift of the position of
the primary star as a function of wavelength. For image o6f109020 
this gives the results shown in Fig. 2.
 As can be seen in Fig. 1, 
the hotter component dominates at short
wavelengths and the cooler star at long wavelengths. This makes 
the relative separations measured at the
ends of the spectra less reliable than those measured at wavelengths 
where the two components have more
comparable fluxes. We therefore derived the adopted stellar separation 
for each image by only using the
median difference between the stellar positions as measured between 
columns 200 and 600, where 
components A (= Aa + Ab) and B were clearly resolved. We estimate that the uncertainty 
in measuring the  projected
separation at a given PA is about 0.03 pixels.  

Results for the separation as a function of position angle as
measured in each image are shown in  Table 1 for the series 
of images taken on June 6, 2002.  A second series of images were 
taken on June 19, 2003.  However, in this series, the Cepheid 
was at a much brighter phase, and measurements were  significantly
less accurate.
%  Note that for the 
%images of W-SGR taken at PA of -74 and
%-45 degrees, the projected separation was two small to measure.  
Combining the measured separations in Table 1, and
assuming that the cross dispersion plate scale for the filtered 
F25NDQ1 G230L image ($0\farcs0248$/pixel)  we find that the
hotter companion for W Sgr is located at a distance of  
$0\farcs1645$ $\pm$ $0\farcs0006$ at a position angle of
210.0 $\pm$ 0.5 degrees.  (Further detail is provided by 
Fig. 10 in Massa and Evans, 2008.)

\subsection{Spectral Extraction}

In order to obtain further information about the companion in the
spectroscopic orbit, the two resolved spectra in the image were extracted. For
this purpose, a STIS observation where the Cepheid is close to minimum light
was selected to enhance the contribution from the close companion. 
The image used (o6f109010) also has 
one of the  the best projected separations (o6f109010 for W Sgr).
The exposure was taken at (mid exposure): 2002 06 06 10:56:00 =
2,452,431.955. Using the period from Szabados (1989)
  P = 7.594904 E =2443374.622, the Cepheid was at phase 0.55
during the observation.  Using the light curves of Moffett and 
Barnes (1984), corrected to this period, the Cepheid parameters at 
this phase are V= 4.92 mag and B-V = 0.96.
The reddening for W Sgr is E(B-V) = 0.11 corrected for the effect
of the hottest companion (Evans, 1991).

We prepared a flat fielded image in which we subtracted out the 
contributions of the hottest star, W Sgr B, using
the best fit parameters derived above. We then used 
the standard STIS STSDAS x1d task to extract the spectrum for
the W Sgr A = Aa + Ab.  To minimize the residual contamination from 
errors in the subtraction of
adjacent stars we used a 3 pixel high extraction box 
to measure the flux. The x1d task automatically corrects
the flux for the encircled energy extraction as a function 
of extraction box size. This gives us an extracted
spectrum for each star while minimizing the contribution for adjacent stars.

\subsection{Comparisons}

As discussed above
 STIS provides  two spatially resolved spectra,
one of the hottest star in the system, the other the composite 
of the Cepheid and the companion in the spectroscopic
orbit.  The next step is to investigate whether we can 
identify the spectroscopic companion in the composite 
spectrum at any wavelength.  In order to do this, 
we have compared the composite spectrum with two 
nonvariable supergiants which bracket the Cepheid in 
(B-V)$_0$ using IUE spectra.  Table 2 lists the relevant parameters for 
the supergiants $\beta$ Aqr and $\alpha$ Aqr.   
The supergiants have been scaled to match the flux 
of the W Sgr spectrum between 2600 and 3000 \AA.

Figs. 3-6 show the comparisons.  The normalized energy
distributions of both $\beta$ Aqr and $\alpha$ Aqr
match that of W Sgr reasonably well from 3200 to 2500 \AA.
Since they bracket W Sgr at this phase in (B-V)$_0$
(Table 2) the expectation is that the normalized spectrum 
of $\beta$ Aqr (G0 Ib) should have a little more flux from
2200 to 1800 \AA\/ and  $\alpha$ Aqr (G2 Ib) should have a little less.
Figs. 4 and 6 show that on the contrary, both supergiants
have more flux at 2200 \AA, and at least a little more 
flux at 1900 \AA.  

We have even arbitrarily scaled the $\alpha$ Aqr spectrum
by 0.85 from 1800 to 2400 \AA\/ (Fig. 7).  Even in this 
comparison there is no indication that the W Sgr flux at 
1800--1900 \AA\/ is any larger than would be predicted 
by the supergiant spectrum.  That is, there is no 
indication of a higher proportion of flux contributed at the 
shortest wavelength due to a companion hotter than the 
Cepheid.

We attribute the extra flux in the supergiants in the 
1800 to 2400 \AA\/ region to a subtle difference found in 
the spectral energy distributions of supergiants covering 
a range of spectral types from F2 Ib to G8 Ib (Evans and Teays, 1996).  They were 
compared with energy distributions of $\delta$ Cep at 
5 phases.  $\delta$ Cep has a period very 
similar to that of W Sgr (5.37 d)
and should have very similar pulsation.  The supergiants were 
found to have more flux than $\delta$ Cep when compared 
at phases with the same (B-V)$_0$.  Morossi et al. (1993)
found a similar result of extra flux at short wavelengths
when comparing a sample of G and K giants with radiative 
Kurucz models.  They attribute this to a non-radiative 
heating in the giants.

Unfortunately we were 
unable to find an IUE exposure in the 1800 to 2000 \AA\/ region 
of a non-binary Cepheid with as cool a value of (B-V)$_0$ as
this phase of W Sgr.  Such a spectrum would presumably 
match the W Sgr spectrum over the full 1800 to 3200 \AA\/
spectral range.  

The difference in energy distributions between W Sgr and 
the supergiants  $\beta$ Aqr and $\alpha$ Aqr limits the 
detection of a companion which is fainter and only 
somewhat hotter than the Cepheid.  Clearly Fig. 3-7 show
no sign of extra flux in the 1800--2000 \AA\/ region from a 
companion hotter than the Cepheid.  
There is one further consideration which limits the information
we can obtain about the companion in the spectroscopic orbit. 
As mentioned above, the configuration used for the W Sgr 
exposures is little used, and not extensively calibrated.  

\subsection{Companion}

Does the W Sgr observation allow us to put any limits
on the spectroscopic companion Ab?  To investigate this, 
we have made comparisons with IUE spectra 
of main sequence stars.  The stars
and their optical properties are listed in Table 3,
together with the IUE spectra used for the comparisons.
Absolute magnitudes are also listed, taken from 
Schmidt-Kaler (1982).   The absolute magnitude for W Sgr 
was taken from the Galactic Cepheid Database (Fernie, Beattie, Evans, 
and Seager:  http://www.astro.utoronto.ca/DDO/research/cepheids/ ).
The mean M$_V$ = -3.76 mag is corrected to -3.51 mag for the phase 
of observation.  Combining this with the information in Table 3, 
the main sequence comparison spectra were scaled as 
appropriate for companions of W Sgr.  The scaled main sequence 
stars and the W Sgr spectra are shown in Figs. 8 and 9.
(Rescaling the companions to the distance found directly by 
Benedict, et al. makes an imperceptible difference to the figures.)

An F0 V companion could be  part of the composite STIS spectrum
of W Sgr Aa + Ab only if there were no flux 
contributed by the Cepheid at 1800-1900 \AA.
The supergiants bracketting W Sgr at the phase of observation  indicate 
that the Cepheid should contribute to the spectrum, at least at 
the level of 0.2 x 10$^{-13}$ ergs cm$^{-2}$ sec$^{-1}$ \AA$^{-1}$.
Even if the supergiant flux overestimates the contribution of the 
Cepheid, as found by Evans and Teays (1996), the flux from Ab would 
still be less than the F0 V star.  
An F5 V companion, on the other hand would have only a minute 
effect on the energy distribution, and could easily be part of the system.

\section{Discussion}

\subsection{The Mass of the Cepheid}

From the discussion in the section above, an F0 V companion is ruled out.
A  companion as cool as F5 V or cooler is possible.  Using the mass-spectral 
 relation from Harmanec (1988), an F0 V and F5 V stars have a masses 
 of 1.5 M$\sun$ and 1.3 M$\sun$ respectively.  We adopt a mass $\leq$ 1.4 M$\sun$
 for the companion.  From the orbital solution of Benedict, et al. (2007),
 the mass of the Cepheid becomes M  $\leq$ 5.4 M$\sun$.

\subsection{Distant Companions} 

The ADS lists two distant possible companions to W Sgr = ADS 11029 = CD -29 14447
= WDS 18050-2935 (Table 4).  We examine the information about these two stars
to see whether they are likely physical companions.  
A recent addition to this information is photometry from the 2MASS point 
source catalogue (Cutri, R., et al. 2003), which is listed in Table 4 for B and C.
The 2MASS photometry for both stars is quality A, except for J for star
C which is an upper limit.
For the brighter Cepheid itself, photometry at mean light has been taken 
from the recent compilation of van Leeuwen, et al. (2007).  This has been transformed 
back from the SAAO photometry to the 2MASS system using the relations provided
by Carpenter (2001).  These colors can be corrected for extinction using the 
E(B-V) = 0.11 mag (Evans, 1991) and an  extinction law of  
Mathis (1979).

  The check we want to make is whether the colors and magnitudes are 
  appropriate for main sequence stars (slightly pre-main sequence 
stars) if they are at the distance of W Sgr.  We use a distance of 
439 pc for W Sgr from the HST Fine Guidance Sensor (FGS) parallax
from Benedict, et al. 2007.  At this distance, components B and C 
have absolute magnitudes M$_K$ of 1.32 and 3.14 mag respectively.  
For comparison, we use the colors and magnitudes for Kraus and 
Hillenbrand (2007b, Table 5).  The unreddened (J-K)$_0$ 
and (H-K)$_0$ for B
corresponds to spectral type from late K to early M in this table.
This range  has M$_K$
$\simeq$ 5.0 mag.  For component C, (H-K)$_0$ and hence  M$_K$ is similar.
Clearly, the observed magnitudes are too bright to be at the 
distance of W Sgr, and hence not physical companions.  

\subsection{Components}

The ADS 11029 system, then, is made up of three components.
The Cepheid (Aa) and the cool companion (Ab) and the hot 
component (B).  If we use the maximum masses (above), 
the mass ratio of this pair is 0.26.  These
masses and the orbital period (Benedict et al.)
correspond to a semimajor axis of  5.0 AU for the Aa--Ab pair,
corresponding to $0\farcs011$.

We note that the absolute magnitude M$_V$ = -3.97 mag 
found by Benedict, et al. from the HST FGS measurements is
in very good agreement with the absolute magnitude found 
from the IUE spectrum of the companion M$_V$ = -4.00 mag
(Evans, 1991).

The third component B, the hottest star, is $\ge$ $0\farcs1645$ 
from A, which corresponds to a distance of $\ge$ 72 AU
at the distance of the system.  This companion might have
been the one found by Morgan, et al., although the 
magnitude difference between it and the Cepheid 
(5.3 mag at V; Evans, 1991) is very challenging.  
This star has a spectral type of A0 V, corresponding
to a mass of 2.2 M$\sun$.  

The ratio of the separations between the outer and inner 
systems is comfortably within Tokovinin's (2004) empirical 
findings for triple systems.

The two more distant stars do not seem to be physically associated with the 
Cepheid system, making a total of 3 stars in the system.

\section{Summary}

The STIS spectrum of W Sgr elucidates some of the puzzles about 
this well studied Cepheid multiple system.  The hottest star 
in the system, studied in IUE spectra is not the companion in the 
low amplitude spectroscopic binary.  It is a more distant third 
member of the system.  

We have compared the ultraviolet spectrum of the unresolved 
spectroscopic binary with supergiant and main sequence spectra 
to put limits on the temperature and mass
of the companion.  A companion as hot as F0 V is not compatible
with the spectrum.  Using this a an upper limit, we infer an 
upper limit to the mass of the companion of 1.4 M$\sun$.  
Combining this with the orbit of the system from Benedict, et al. (2007),
we find a mass for the Cepheid of $\leq$ 5.4  M$\sun$.

We have also investigated two more distant possible companions to 
see whether they are likely to be additional members of the system.
The absolute magnitudes and colors were calculated from  2MASS J, H, and K
magnitudes assuming the two stars are the distance of the W Sgr system.
The resulting magnitude-color combination is not consistent with 
main sequence stars.  That is, the two stars at 33" and 46" are not
likely to be companions, and the system is limited to three components.

%% The equation environment will produce a numbered display equation.

%% The \notetoeditor{TEXT} command allows the author to communicate
%% information to the copy editor.  This information will appear as a
%% footnote on the printed copy for the manuscript style file.  Nothing will
%% appear on the printed copy if the preprint or
%% preprint2 style files are used.

%% The eqnarray environment produces multi-line display math. The end of
%% each line is marked with a \\. Lines will be numbered unless the \\
%% is preceded by a \nonumber command.
%% Alignment points are marked by ampersands (&). There should be two
%% ampersands (&) per line.

%\section{Floating material and so forth}

%% The displaymath environment will produce the same sort of equation as
%% the equation environment, except that the equation will not be numbered
%% by LaTeX.


%% Included in this acknowledgments section are examples of the
%% AASTeX hypertext markup commands. Use \url without the optional [HREF]
%% argument when you want to print the url directly in the text. Otherwise,
%% use either \url or \anchor, with the HREF as the first argument and the
%% text to be printed in the second.

\acknowledgments

We are happy to acknowledge valuable conversations with Brian Mason about
double stars.  
We would like to thank Karla Peterson of STScI for valuable 
guidance in the preparation of the observations.
Financial support was provided by Chandra X-ray  Center NASA Contract NAS8-03060 (for NRE) 
Support from the research program for CSC astronomers                                
at STScI under contract STI-111271 is gratefully acknowledged.                       
Support for HST Program number 9105 was provided by NASA through a                   
grant from the Space Telescope Science Institute, which is operated                  
by the Association of Universities for Research in                                   
Astronomy, Inc., (AURA), for the National Aeronautics and Space                      
Administration (NASA) under Contract NAS5-26555.
This research made use of This research has made use of the SIMBAD database,
operated at CDS, Strasbourg, France, and of the IUE reduction software 
package.

\clearpage

\begin{deluxetable}{lrc}
%\tabletypesize{\scriptsize}
\tablecaption{  Separation of Components A and B }
%\label{tbl-1}}
\tablewidth{0pt}
\tablehead{
\colhead{Observation} & \colhead{PA} & \colhead{Sep} \\
\colhead{} & \colhead{deg} & \colhead{pix} }

\startdata

o6f107010 & -115.0  &   5.441   \\
o6f107020 & -115.0  &   5.443   \\ 
o6f107030 & -115.0  &   5.470   \\ 
o6f108010 & -86.0   &   2.923   \\ 
o6f108020 & -86.0   &  2.919   \\ 
o6f108030 & -86.0    &  2.914   \\ 
o6f109010 & -144.0 &  6.614   \\ 
o6f109020 & -144.0 &  6.612   \\ 
o6f109030 & -144.0 &  6.554   \\

 \enddata

\end{deluxetable}        

\begin{deluxetable}{lrccc}
%\tabletypesize{\scriptsize}
\tablecaption{Supergiant Comparisons }
%\label{tbl-1}}
\tablewidth{0pt}
\tablehead{
\colhead{Star} & \colhead{E(B-V)} & \colhead{(B-V)$_0$} & \colhead{IUE Long} &
\colhead{IUE Short}   \\
\colhead{} & \colhead{mag} & \colhead{mag} & \colhead{} &
\colhead{}}

\startdata

W Sgr  &  0.11 & 0.85 & & \\
$\beta$ Aqr & 0.03 &  0.81 & LWR06031 &  SWP07124 \\
$\alpha$ Aqr & 0.07 & 0.91 & LWR12113 & SWP01623  \\

 \enddata

\end{deluxetable}

\begin{deluxetable}{lrcc}
%\tabletypesize{\scriptsize}
\tablecaption{Main Sequence Comparisons }
%\label{tbl-1}}
\tablewidth{0pt}
\tablehead{
\colhead{} & \colhead{A0 V} & \colhead{F0 V} & \colhead{F5 V} \\ }

\startdata

Star  &     HD 103287  & HD 12311 &  HD 27524 \\
V (mag)  &     2.44  &   2.86  &    6.80 \\     
M$_V$ (mag)  &      1.3  &    2.8 &      3.6 \\
E(B-V) (mag) &  0.01  &   0.00  &      0.00 \\
IUE Long   &   LWR07124 & LWR 09862 & LWR12183  \\ 
IUE Short   & SWP08198 & SWP11242 & SWP15819 \\
                                                                           
 \enddata

\end{deluxetable}

\begin{deluxetable}{lrccc}
%\tabletypesize{\scriptsize}
\tablecaption{Distant Companions }
%\label{tbl-1}}
\tablewidth{0pt}
\tablehead{
\colhead{Star} & \colhead{Sep } & \colhead{K } & \colhead{J-K}  & \colhead{H-K}  \\ 
\colhead{ } & \colhead{ "} & \colhead{ mag} & \colhead{mag}  & \colhead{mag}  \\ }
\startdata

W Sgr =  ADS  11029 A  &  --  & 2.83  & 0.46  & 0.08  \\  
 ADS  11029 B         &  33  &  9.57  & 0.61  & 0.24  \\
 ADS  11029 C         &  46 &  11.39  & --  & 0.20
                                                                           
 \enddata

\end{deluxetable}

%% Text for table notes should follow after the \enddata but before
%% the \end{deluxetable}. Make sure there is at least one \tablenotemark
%% in the table for each \tablenotetext.

%\tablenotetext{a}{Sample footnote for table~\ref{tbl-1} that was generated
%with the deluxetable environment}
%\tablenotetext{b}{Another sample footnote for table~\ref{tbl-1}}

%\tablecomments{Occasionally, authors wish to append a short
%paragraph of explanatory notes that pertain to the entire table, but
%which are different than the caption.  Such notes should be placed in
%a {\tt tablecomments} command like this.}

%% If you use the table environment, please indicate horizontal rules using
%% Do not put multiple tabular environments within a single table.
%% The optional \label should appear inside the \caption command.

\clearpage   

%\figcaption[]{The STIS image O6NE20010.  Wavelength increases from left to right.  The upper spectrum
%is the Cepheid (Aa) plus the cool companion (Ab); the lower spectrum is component B. 
%The intensity is shown on a log scale. The cross-dispersion has  been stretched
%by a factor of 8 to emphasize the resolved components. The wavelength covers approximately
%1800 to 3200 \AA.  The 2800 \AA\/ line in the Cepheid spectrum is the most prominent
%feature (to the right of center.) 
%(Enjoy the full color image on the electronic version.)
%   \label{fig1}} 

\figcaption[]{A flat-fielded image   o6f109020 of W Sgr.  
 Wavelength increases to the right in this figure from about 
  1796 to 3382 \AA. The cooler
component is the upper one in the image, and clearly separated from 
the hotter component below which extends much further toward 
short wavelengths.  The spectrum has been compressed by a factor of 8 in 
the horizontal direction and is  on a log scale.
The strongest feature in the Cepheid spectrum is 
2800 \AA.
   \label{fig1}}

\figcaption[]{The separation between Aa + Ab (cool component)  
and B (hot component).  
 Measured position is shown as a function of wavelength for the two components of
W Sgr as seen in observation o6f109020.  
The position of the spectral trace has been adjusted
to remove any systematic shift of the cool component as a function of wavelength. 
   \label{fig2}}

\figcaption[]{W Sgr A (solid) compared with $\beta$ Aqr (dashed).
$\beta$ Aqr has been more heavily smoothed than W Sgr since 
the IUE camera from 2000 to 2400 \AA\/ is very noisy. In all spectral
figures,  wavelength is in \AA, flux is in ergs cm$^{-2}$ sec$^{-1}$ \AA$^{-1}$.  
   \label{fig3}}                                   

\figcaption[]{The same spectra as in Fig. 3.  The scale has been increased to 
focus on the comparison at the shortest wavelengths of the STIS W Sgr spectrum
   \label{fig4}}                                   

\figcaption[]{W Sgr A (solid) compared with $\alpha$ Aqr (dashed).
Again,  $\alpha$ Aqr is more heavily smoothed than W Sgr.  
   \label{fig5}}                                   

\figcaption[]{The same spectra as in Fig. 5, emphasizing the short
wavelength region of W Sgr.
   \label{fig6}}                                   

\figcaption[]{The same spectra as in Fig. 6, except that the flux of 
$\alpha$ Aqr has been arbitrarily reduced by 15\%.
   \label{fig7}}

\figcaption[]{W Sgr (solid) compared with the A0V, F0 V, and F5 V  stars 
 (dashed) from top to bottom.  The F0 V and F5 V stars are only barely 
 distinguishable at the longer wavelengths.
   \label{fig8}}

\figcaption[]{The same spectra as in Fig. 8 with flux and wavelength scales
adjusted to emphasize the 1800--2000 \AA\/ region. Only the F0 V (upper) and F5 V
spectra are visible;  the A0 V spectrum is too bright.   
   \label{fig9}}

%\figcaption[]{Images of white dwarf calibration stars observed with ACS-HRC
%the F220W filter. No artifact is seen in the location of the Polaris
%companion. From left to right the stars with date of observation are: GD71
%(2002/4/12), G191B2B (2003/8/30), GD 71 (2004/2/3), and GD 153 (2005/5/19
%\label{fig3}} l l 
   
%\figcaption[]{ U Aql rectified
%   \label{fig2}} 
   
% plots in /data/cygnus/hst/massa/multip/proffitt
\clearpage                  
 
%\plotone{wsgr.stis.ps}  
 
%\clearpage      

%\plotone{uaql.ps}  
 
%\plotone{o6f109020_detail.ps}

\plotone{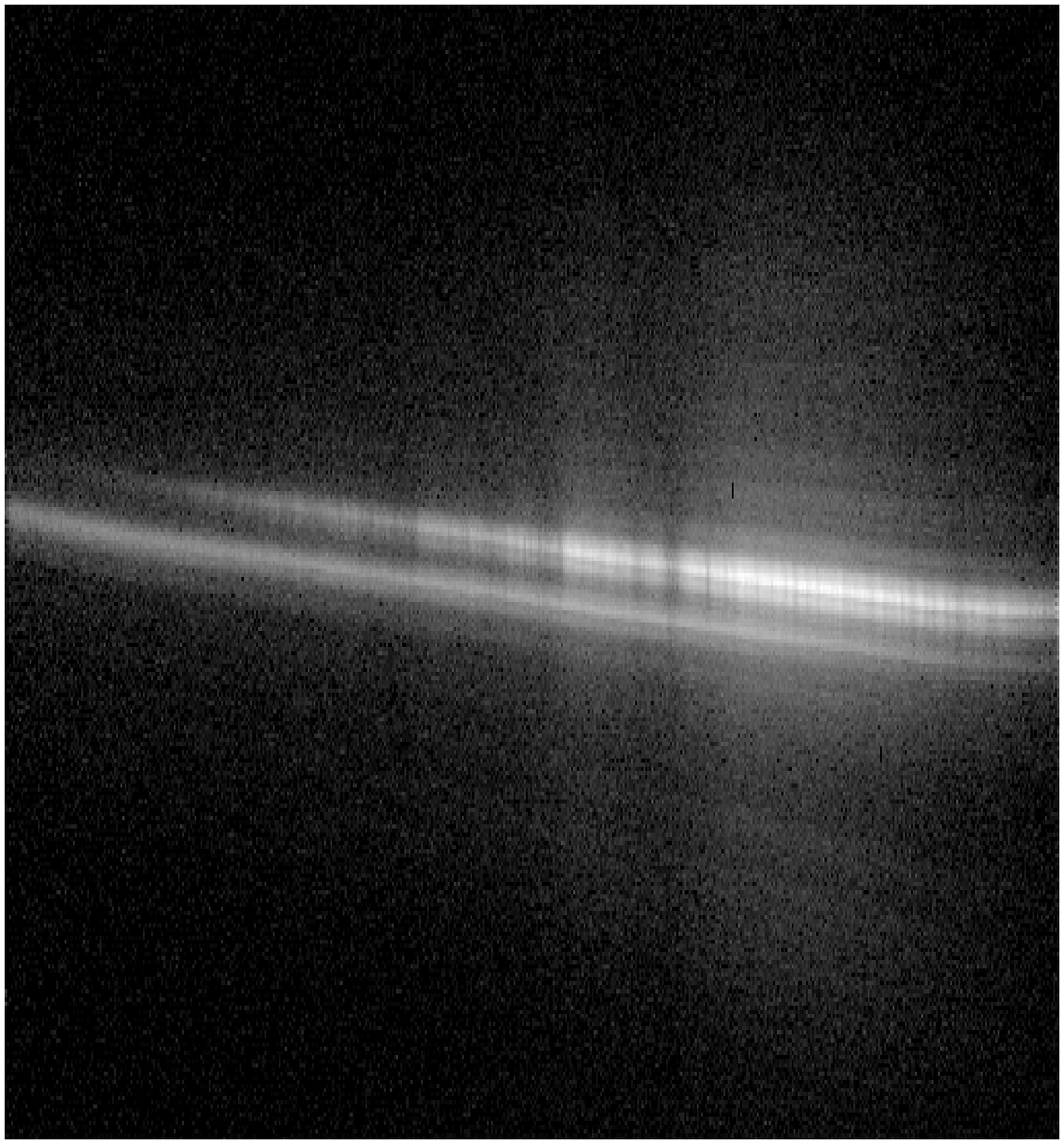}

\plotone{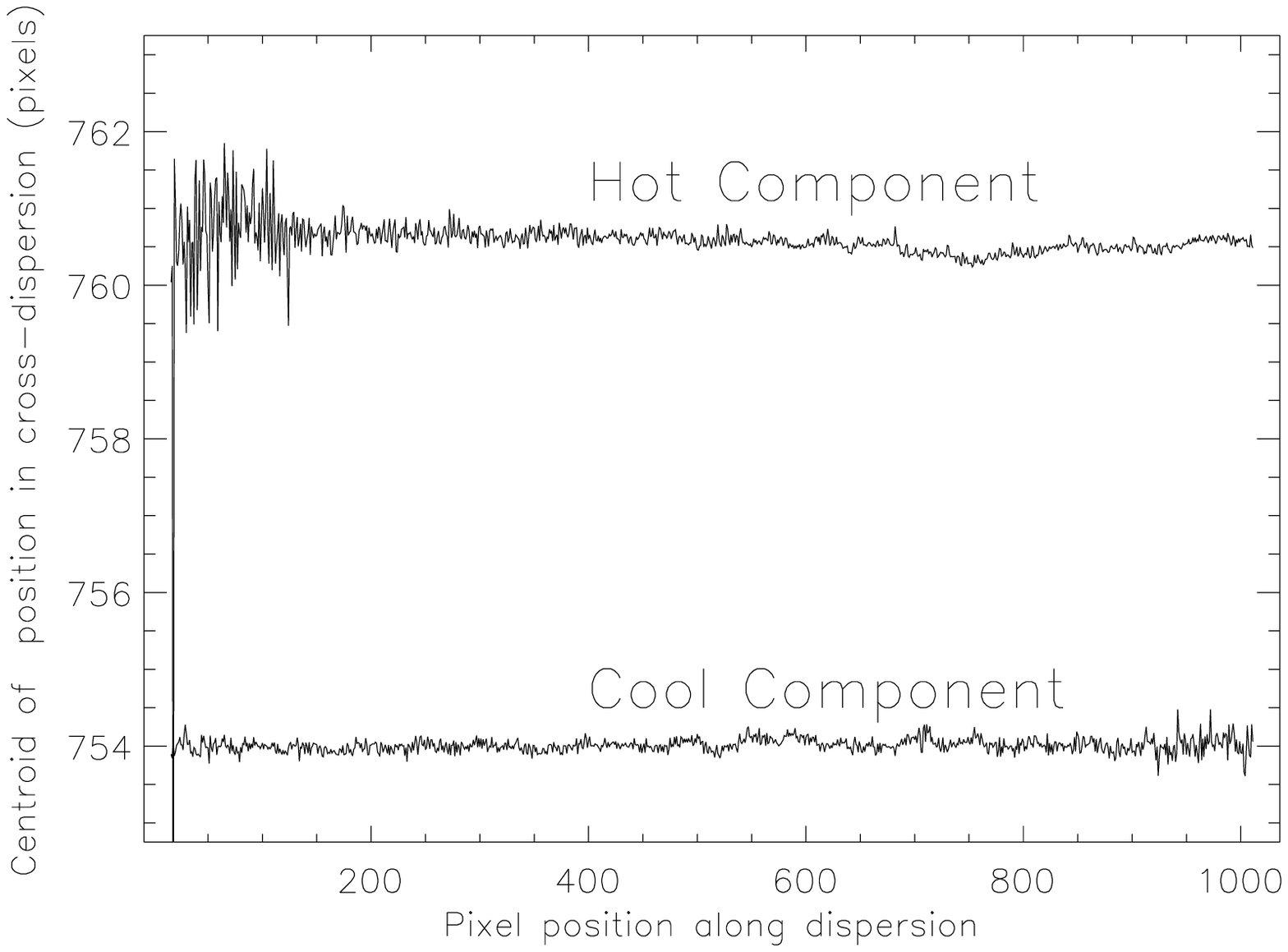}

\plotone{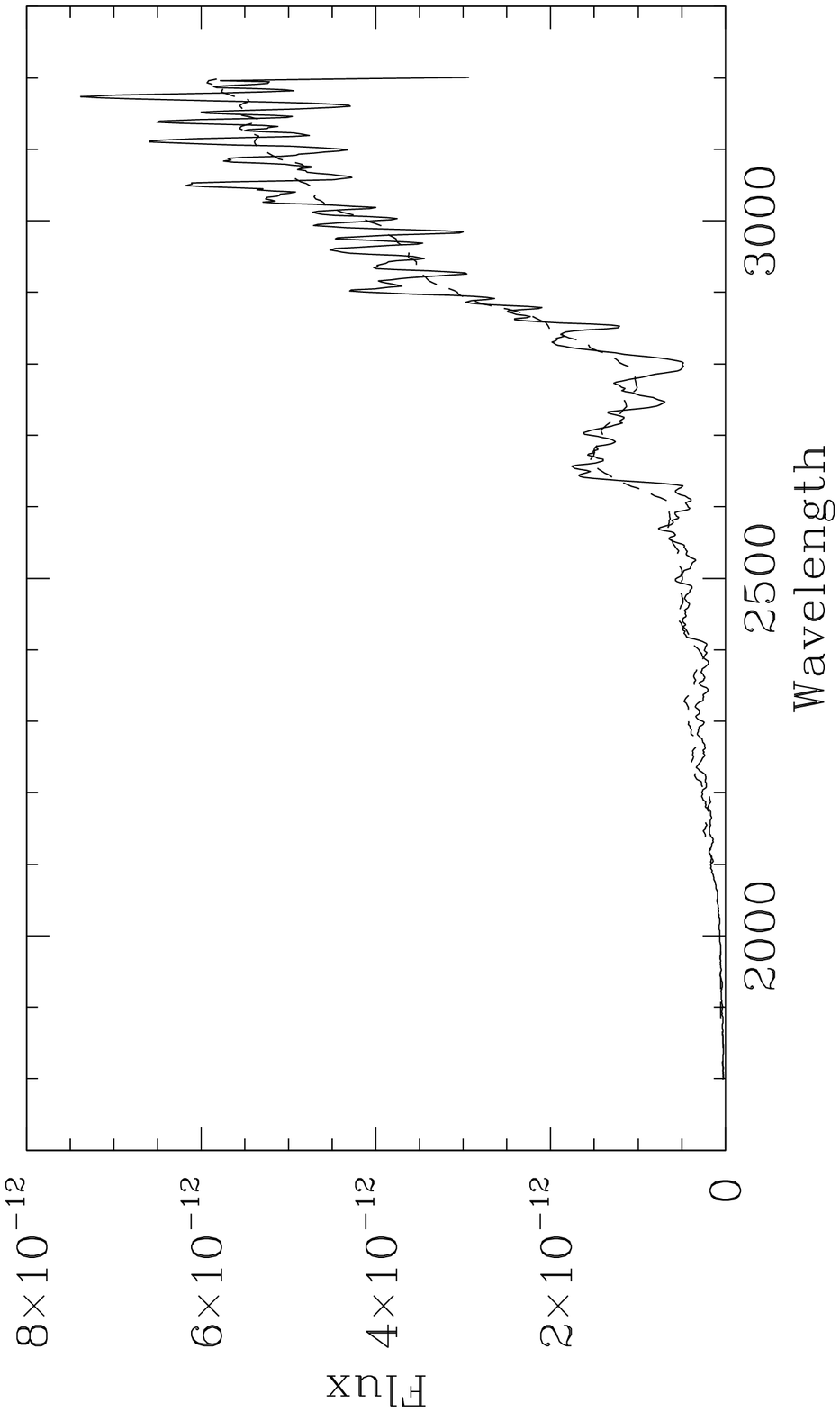}

\plotone{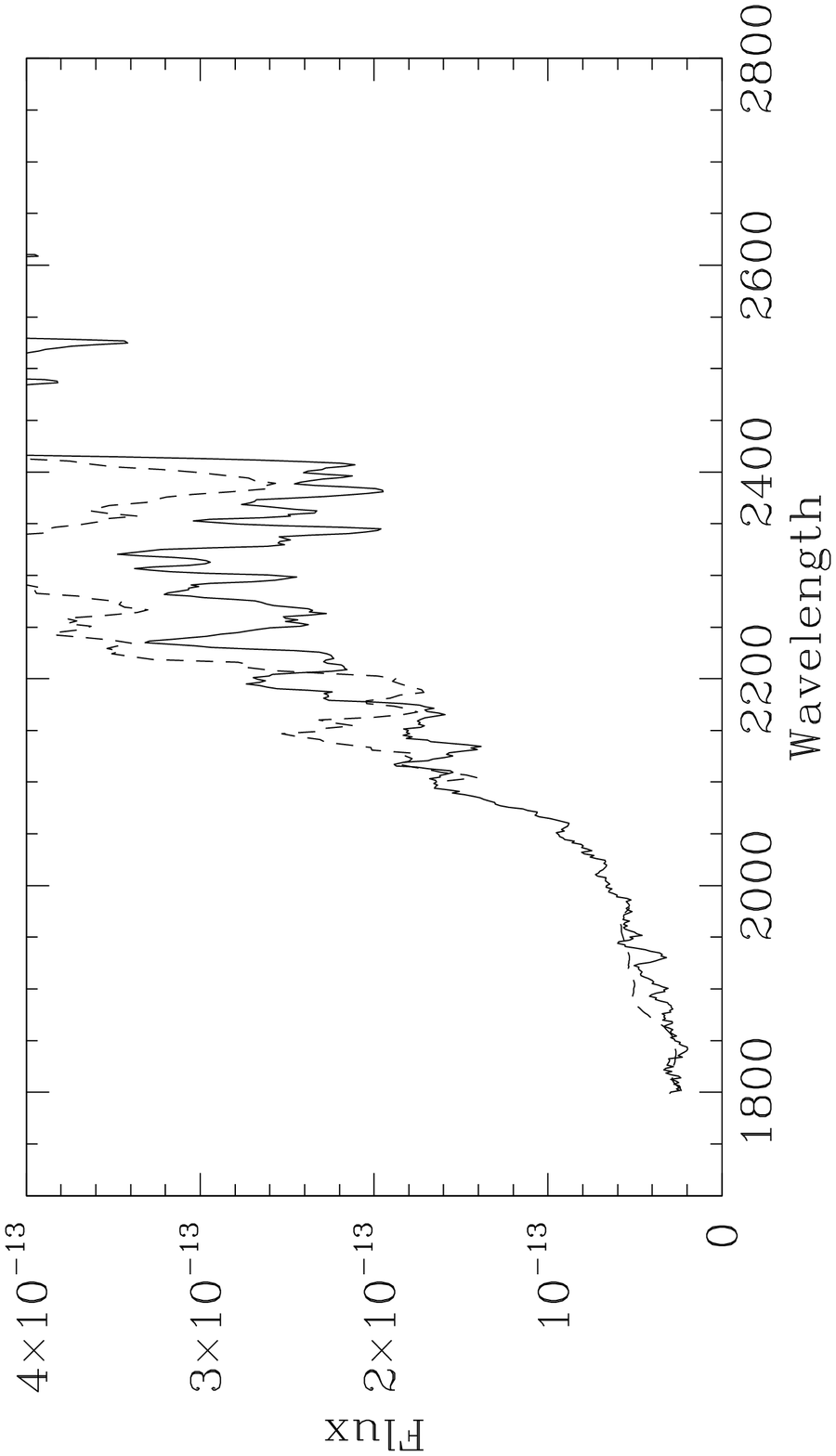}

\plotone{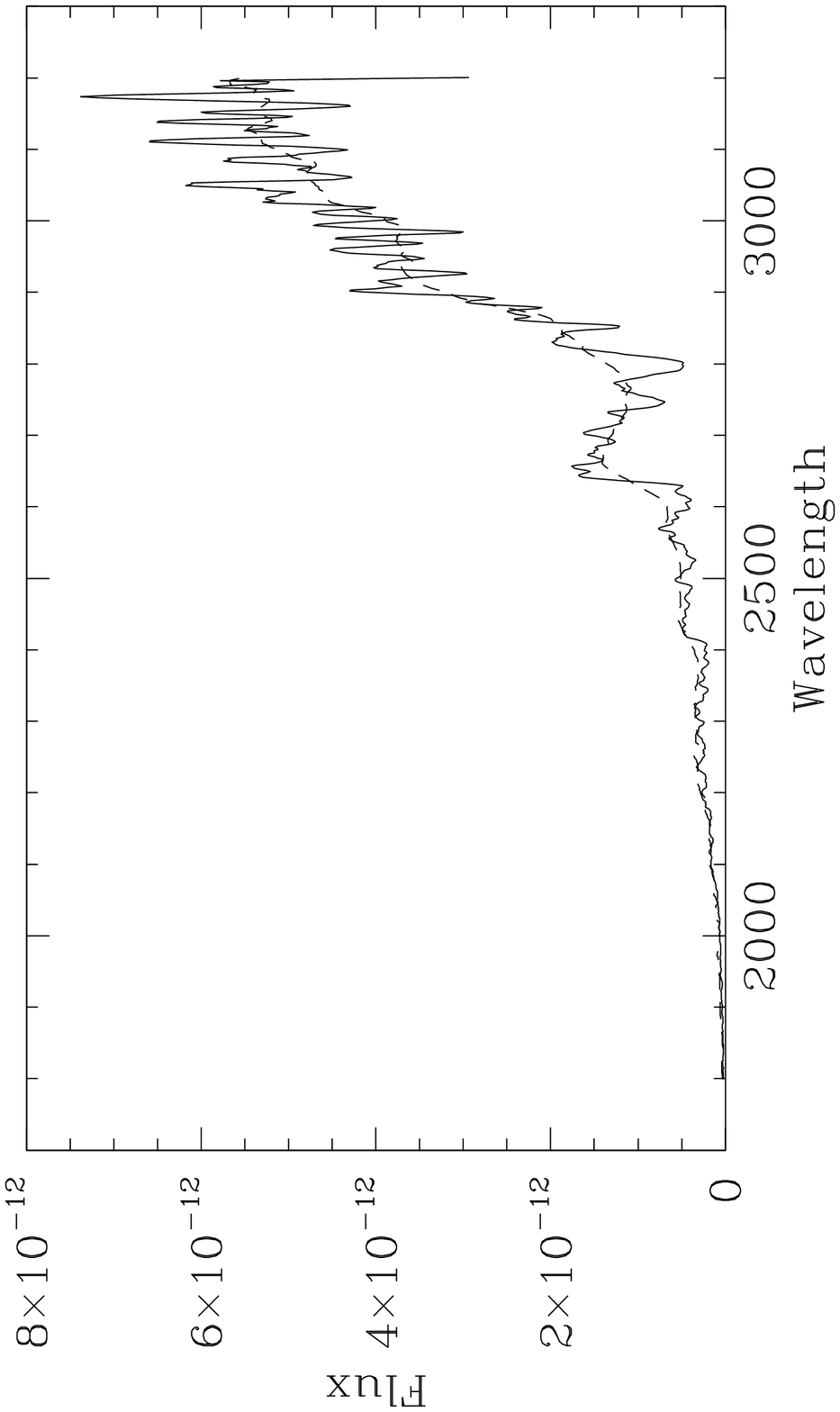}

\plotone{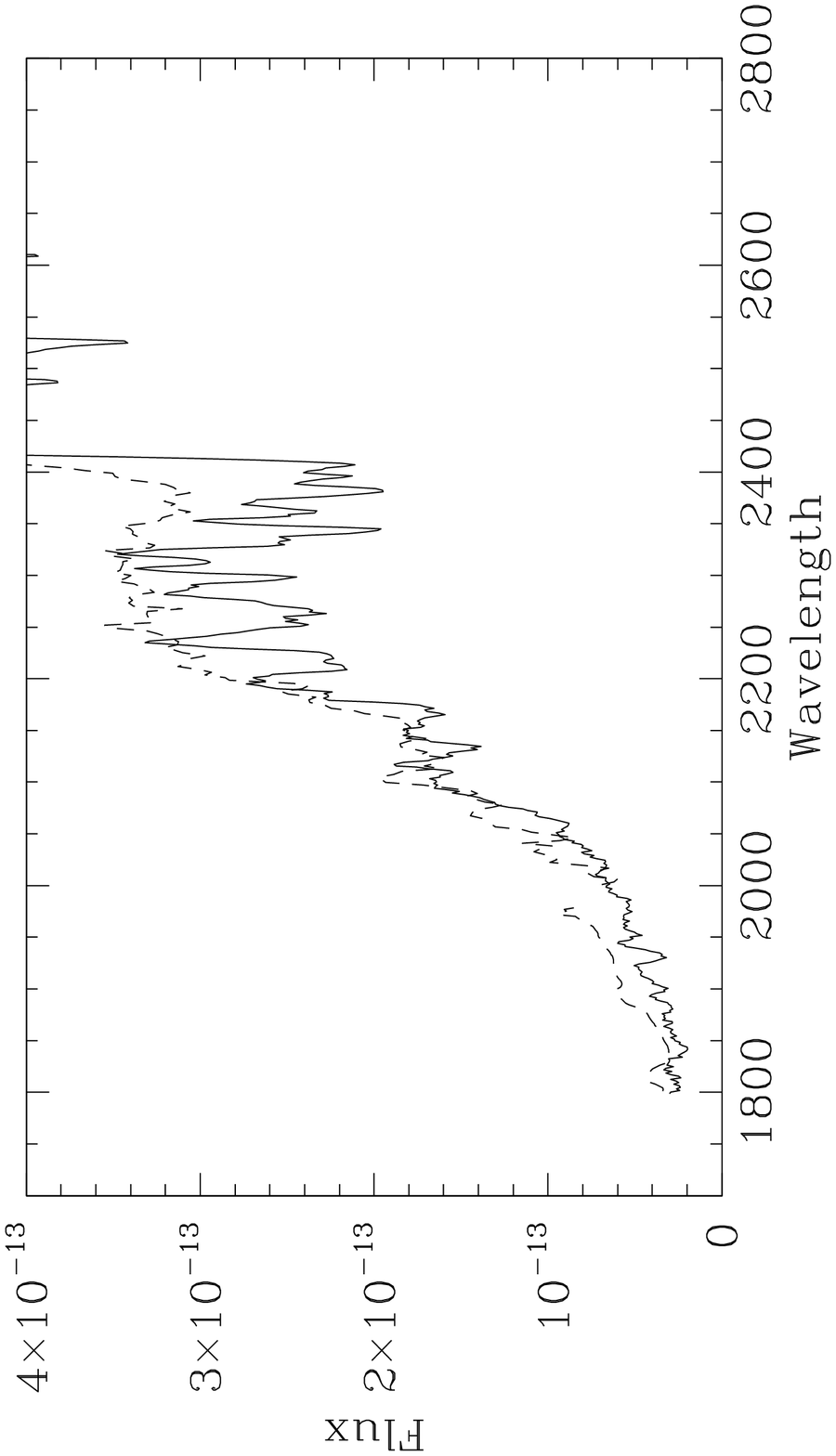}

\plotone{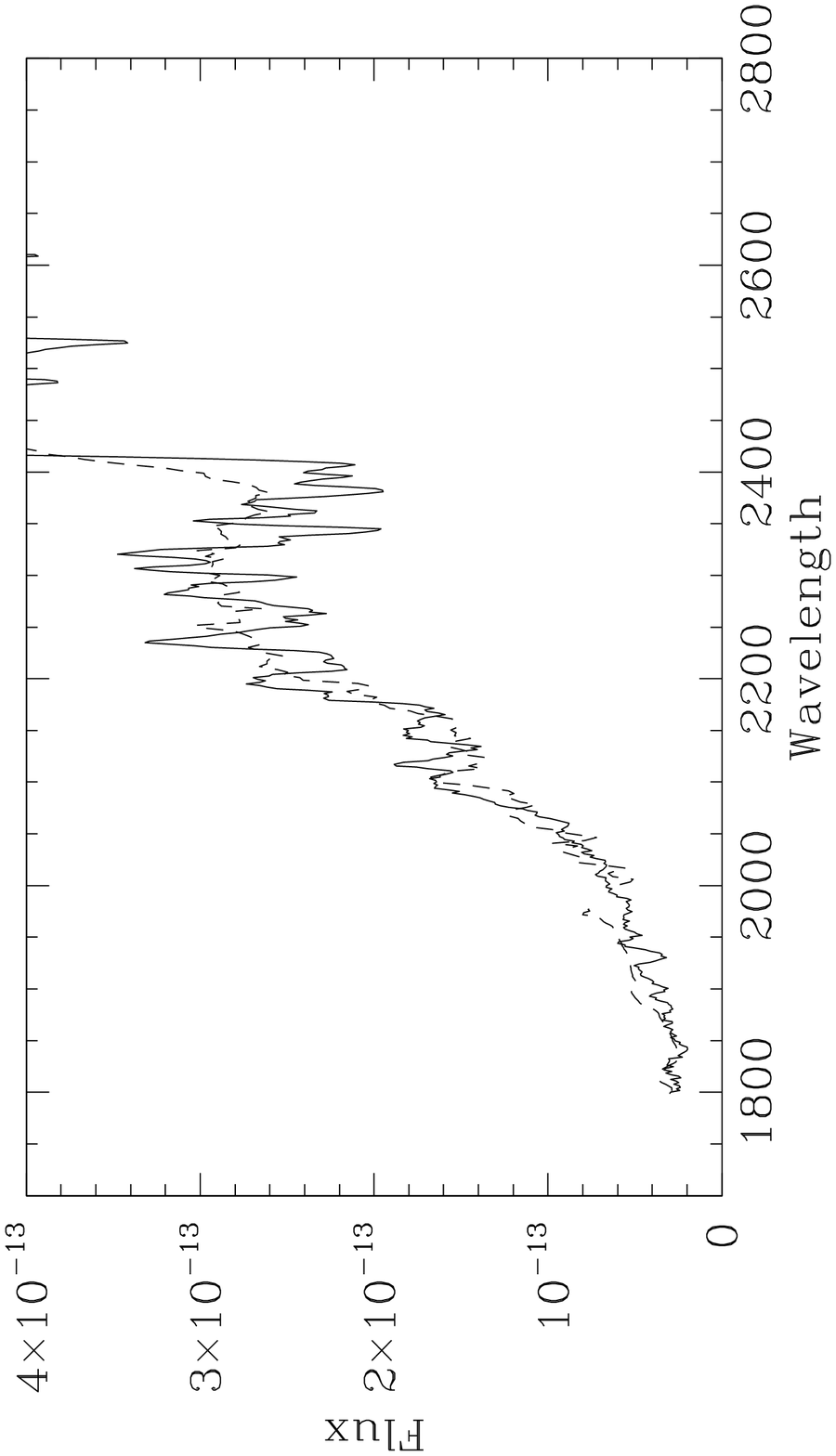}

\plotone{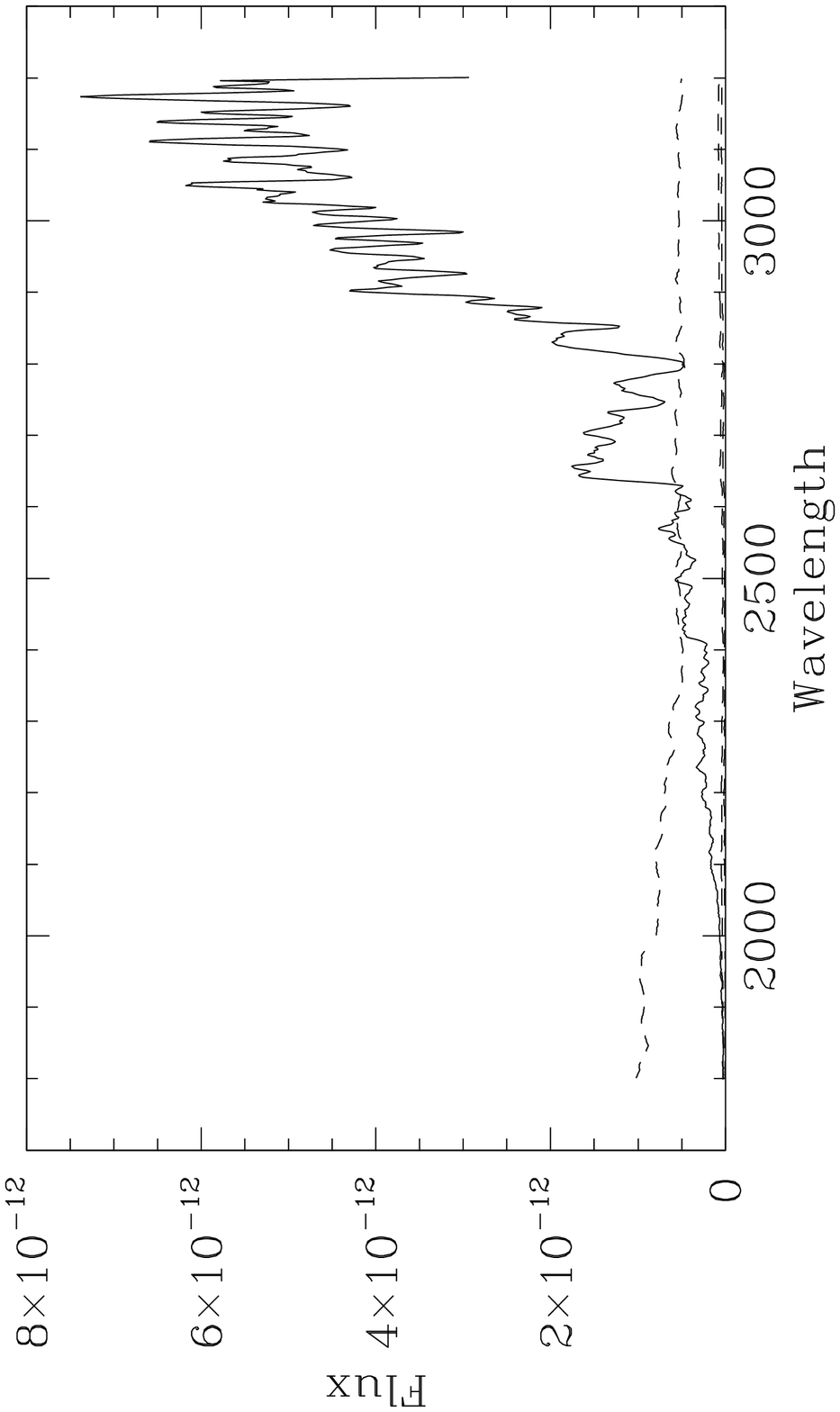}

\plotone{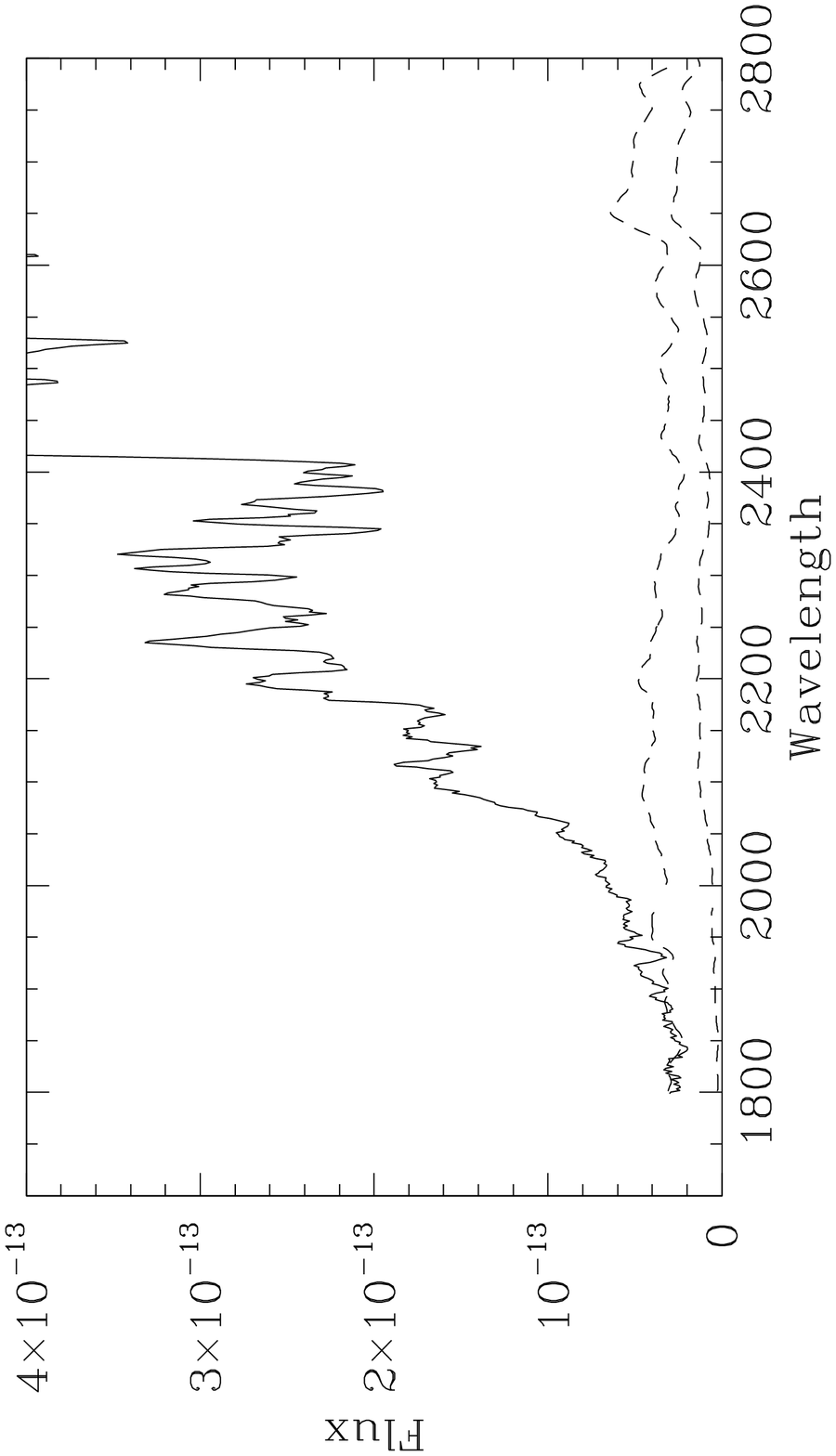}

%% The following command ends your manus\ipt. LaTeX will ignore any text

\end{document}